\documentclass[jeosman,preprint,fleqn,showpacs,showkeys]{revtex4}
\usepackage{graphicx}
\usepackage{amssymb,amsmath,bm}

\begin{document}

\title{Conditional Generation Scheme for Entangled Vacuum Evacuated Coherent States
by Mixing Two Coherent Beams with a Squeezed Vacuum State}

\author{Sun-Hyun Youn\footnote{email: sunyoun@chonnam.ac.kr, fax: +82-62-530-3369}}
\address{Department of Physics, Chonnam National University, Gwangju 500-757, Korea}

\begin{abstract}

 Conditions to generate  high-purity entangled
vacuum-evacuated coherent states ($|0>|\alpha>^0 - |-\alpha>^0 |0>$
were studied for two cascade-placed beam splitters, with one
squeezed state input and two coherent state inputs whenever a single
photon is detected. Controlling the amplitudes and the phases of the
beams allows for various amplitudes of the vacuum-evacuated coherent
states ($|\alpha>^ 0 = |\alpha> -e^{-|\alpha|^2} |0> $) up to
$\alpha = 2.160$  to be manipulated with high-purity.



\end{abstract}


\maketitle

\section{Introduction}

Quantum information science has taken advantage of the basic quantum
nature of the world.  Superposition and entanglement are the most
important properties in quantum information science. Superposition
of states makes it possible to build a qubit which is a basic logic
element of the quantum computer. A qubit is a state ($ \alpha |0> +
\beta |1>$) different from the classical bit state $|0>$ or $|1>$.
The superposed quantum state is relatively easy to generate for
small photon numbers, but it's challenging  to generate macroscopic
superposed states such as that of Schrodinger's cat.

  The other quantum nature is entanglement which has no classical
analogy. Twin photon generation by parametric amplification is one
of the good examples. Entangled states at the single photon level
have played huge roles in quantum information science. However,
generating a macroscopic entangled state is very difficult. The
entangled coherent state is one of the macroscopic entangled states.
After Sanders theoretical work on entangled coherent states, several
groups studied entangled coherent states by both theoretical and
experimental methods \cite{Sanders11}.

Using a nonlinear Mach-Zehnder interferometer, a pair of coherent
states might be transformed into an entangled superposition of
coherent states \cite{Sanders92}. However, in order to generate an
entangled superposition of coherent states, an ultra high Kerr
effect should be obtained. A scheme to generate a macroscopic
superposition of coherent states using a beam splitter, homodyne
measurement, and a very small Kerr nonlinear effect was proposed by
H. Jeong \cite{Jeong04}. Furthermore, three mode W-type entangled
coherent states\cite{An04} using a single-photon source and Kerr
nonlinearities was suggested \cite{Jeong06}. K. Park et. al.
proposed schemes to implement the superposition of coherent
displacements and squeezings on two beams of light for non-Gaussian
entanglement ( $|\alpha>_1 |0>_2 + |0>_1 |\alpha>_2 $)
\cite{Park15}. With the implementation of a coherent superposition
of two distinct quantum operations, the following hybrid
entanglement is experimentally demonstrated \cite{Jeong13}
\begin{eqnarray}
\frac{1}{\sqrt{2}}(|0>|\alpha> + |1>|-\alpha>). \label{ECb}
\end{eqnarray}

A macroscopic quantum entangled state is essential for developing
quantum information science. One of the simple applications is for
measuring the phase precisely. Joo et. al. presented an improved
phase estimation scheme employing entangled coherent states and
showed  that entangled coherent states gave the smallest variance in
the phase parameter under perfect and lossy conditions\cite{Joo11}.
Recently a new scheme was proposed for measuring a phase to a
precision significantly better than that attainable by both
unentangled ¡®classical¡¯ states and highly-entangled NOON states
over a wide range of different losses with a realistic and practical
technology \cite{Knott14}.

Our work is to demonstrate how to generate entangled coherent states
 with a squeezed light, two coherent lights, two beam splitters
and single photon counters. In our proposed system, we added two
coherent beams with cascade-placed beam splitters  to the
nonclassical state. The two beam splitters and the two coherent
beams give us a degree of freedom to control the output in a highly
nonclassical manner. We characterize  the two outputs from the three
input beams with a detection of a single photon. Our system has a
great advantage in that it can generate a high-fidelity Fock state
\cite{youn13, youn14, youn15}.
\begin{figure}[htbp]
\centering
\includegraphics[width=15cm]{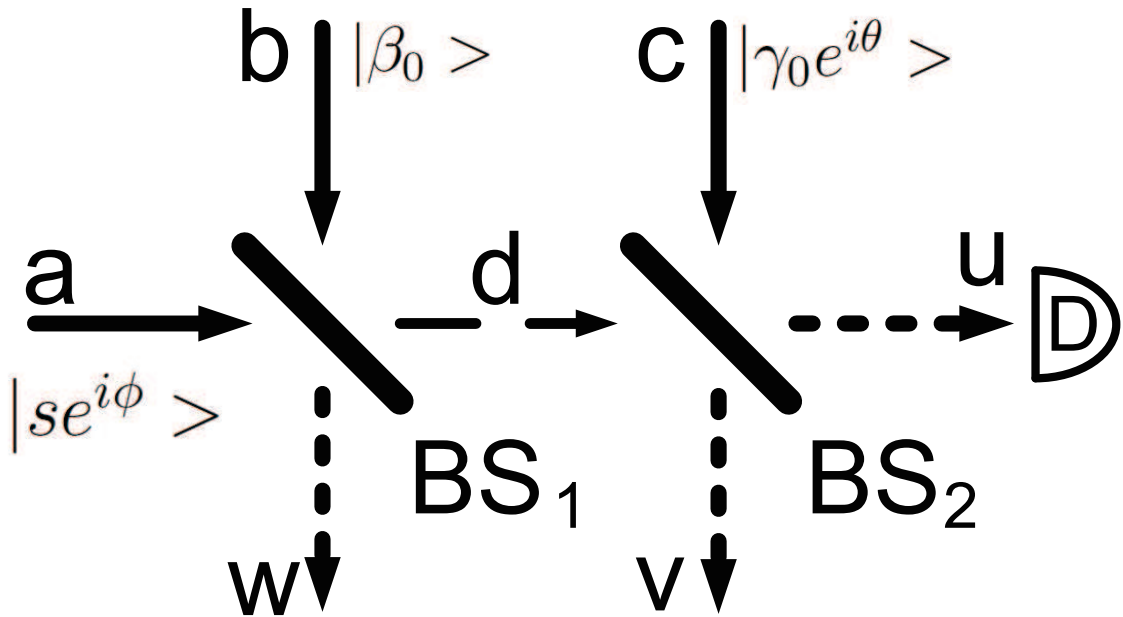}
\caption{Schematic diagram of entangled coherent state generation.
One squeezed state is in the input mode $a$ ($ |s e^{i \phi}> $ ),
and two coherent states ($ |\beta_0 >, |\gamma_0 e^{i \theta}> $)
are in the input modes $b$ and $c$.  BS: Beam Splitter. D:
Detector.} \label{expsetupA}
\end{figure}

  The present paper is organized as follows: In Section II, we
 introduce two-cascade-placed beam splitters with one squeezed state input
  and two coherent state
 inputs. We explicitly calculate the probability of the
  amplitude when a detector  at the output port
 detects a single photon. In section III, we find the analytic condition for
 which an entangled quantum state is generated at
 the output ports $w$ and $v$ when a single photon is detected at the output port $u$.
 In Section IV, we numerically find the conditions for an entangled vacuum
 evacuated coherent state and characterize the entangled state. In
 Section V, we summarize the  main results and discuss the experimental implementation.

\section{Two cascade placed beam splitters}

Let a squeezed vacuum state $|\xi > $ be in the input mode $a$, and
the two coherent states, $|\beta>$ and $|\gamma>$,  be in the input
modes $b$ and $c$, as seen in the experimental set up in Fig.
\ref{expsetupA}.  Then, the input states $|\xi> , |\beta>$, and $
|\gamma>$ can be expressed in the number-state representation
\cite{Loudon}:
\begin{eqnarray}
 |\xi,\beta,\gamma>
  =  e^{-\frac{1}{2} (|\beta|^2+|\gamma|^2 )} \sum_{n = 0, l =0,m =0
  } C_{n}(\xi) \frac{({\hat a}^{\dagger})^n } {\sqrt{n!}} \frac{(\beta {\hat b}^{\dagger})^l } {l!}
  \frac{(\gamma {\hat c}^{\dagger})^m } {m!}  |0>_{a}|0>_{b}|0>_{c}, \label{coherentSt}
\end{eqnarray}
where $C_{n}(\xi)$ is the coefficient of the squeezed vacuum with
squeezing parameter,  $s e^{i \phi} $, and is zero for all odd
values of $n$  and nonzero only for an even value of $n$. The
nonzero values of  $C_{n}(\xi)$ for even values of $n$ become
\cite{Loudon} 
\begin{eqnarray}
  C_{n}(\xi) =  \frac {\sqrt{n !}}{\sqrt{\cosh s} \frac{n}{2}!}
  (- \frac {1}{2} e^{i \phi} \tanh s )^{\frac{n}{2}}.   \label{sV}
\end{eqnarray}

With the experimental set up of Fig. \ref{expsetupA}, the three
creation operators $ {\hat a} ^{\dagger}, {\hat b} ^{\dagger}, {\hat
c} ^{\dagger}$ are written in terms of three creation operators $
{\hat w} ^{\dagger}, {\hat v} ^{\dagger}, {\hat u} ^{\dagger}$ in
output modes $u$, $v$, and $w$. Using an operator relation
\cite{Teich1989}, we can obtain the relationships  between the input
modes and the output modes as the following:
\begin{eqnarray}
 \left(\begin{array}{c}
{\hat a}^\dagger  \\
{\hat b}^\dagger  \\
\end{array}\right)
 = \left(\begin{array}{cc}
 t_1  e^{ -i \phi_{\tau_1} } &  \sqrt{1- {t_1}^2} e^{ -i \phi_{\rho_1} } \\
-\sqrt{1- {t_1} ^2} e^{i \phi_{\rho_1}  } & t_1  e^{i \phi_{\tau_1} } \\
\end{array}\right) \left(\begin{array}{c}
{\hat d}^\dagger  \\
{\hat w}^\dagger  \\ \end{array}\right),  \label{BSab}
\end{eqnarray}
\begin{eqnarray}
 \left(\begin{array}{c}
{\hat d}^\dagger  \\
{\hat c}^\dagger  \\
\end{array}\right)
 = \left(\begin{array}{cc}
 t_2  e^{ -i \phi_{\tau_2} } &  \sqrt{1- {t_2}^2} e^{ -i \phi_{\rho_2} } \\
-\sqrt{1- {t_2} ^2} e^{i \phi_{\rho_2}  } & t_2  e^{i \phi_{\tau_2} } \\
\end{array}\right) \left(\begin{array}{c}
{\hat u}^\dagger  \\
{\hat v}^\dagger  \\ \end{array}\right).  \label{BScd}
\end{eqnarray}

Then the three input creation operators $ {\hat a} ^{\dagger}, {\hat
b} ^{\dagger}, {\hat c} ^{\dagger}$ are written in terms of the
three output creation operators $ {\hat w} ^{\dagger}, {\hat v}
^{\dagger}, {\hat u} ^{\dagger}$ as follows:
\begin{eqnarray}
{\hat a} ^{\dagger} &=& q_{a}^{u}  {\hat u}^{\dagger}+q_{a}^{v}
{\hat v}^{\dagger}+ q_{a}^{w}  {\hat w}^{\dagger} \nonumber \\
 {\hat b}^{\dagger} &=&  q_{b}^{u}  {\hat u}^{\dagger}+q_{b}^{v}  {\hat
v}^{\dagger}+q_{b}^{w}  {\hat w}^{\dagger} \nonumber \\
{\hat c} ^{\dagger} &=& q_{c}^{u}  {\hat u}^{\dagger}+q_{c}^{v}
{\hat v}^{\dagger} ,  \label{opRelations}
\end{eqnarray}
 where $q_{\mu}^{\nu}$ ($\mu= a,b,c, \nu=u,v,w$) represents the
relations between the operators in the input modes (${\hat
a}^\dagger ,{\hat b}^\dagger, {\hat c}^{\dagger}$) and those in the
output modes (${\hat u}^\dagger ,{\hat v}^\dagger, {\hat
w}^{\dagger}$) as follows \cite{youn14}:
\begin{eqnarray}
\{ q_{a}^{u} , q_{a}^{v} , q_{a}^{w}  \} &=& \{  e^{- i (
\phi_{\tau_1}+\phi_{\tau_2} )} t_1
 t_2 ,  e^{- i ( \phi_{\rho_2}+ \phi_{\tau_1})} t_1 \sqrt{1-{t_2}
^2},  e^{-i  \phi_{\rho_1}} \sqrt{1-{t_1} ^2}  \}, \nonumber \\
\{ q_{b}^{u}, q_{b}^{v}, q_{b}^{w} \} &=& \{ e^{ i \phi_{\rho_1}}
\sqrt{1-{t_1} ^2}  e^{ -i \phi_{\tau_2}} t_2, - e^{ i \phi_{\rho_1}}
\sqrt{1-{t_1} ^2}  e^{-i \phi_{\rho_2}} \sqrt{1-{t_2} ^2} , e^{i
\phi_{\tau_1}} t_1 \}, \nonumber \\
\{  q_{c}^{u} ,q_{c}^{v} \} &=&  \{  -e^{ i \phi_{\rho_2}}
\sqrt{1-{t_2} ^2},  e^{i  \phi_{\tau_2}} t_2 \} \label{eqOPabc}.
\end{eqnarray}

 Then, the input states in Eq. \ref{coherentSt} can be written as
number-state representations of the output modes ($u, v, w$) as
follows \cite{youn13}:
\begin{eqnarray}
 |\xi,\beta,\gamma>
  &=& e^{-\frac{1}{2} (|\beta|^2+|\gamma|^2 )} \sum_{n' = 0, l' =0, m' =0} C_{n} (\xi)\beta^{l}\gamma^{m} \frac{\sqrt{n!}}{n_u ! n_v ! n_w !}
   \frac{1}{l_u ! l_v ! l_w !}
\frac{1}{m_u ! m_v! }  \nonumber \\
   & \times &
(q_{a}^{u})^{n_u}  (q_{a}^{v})^{n_v} (q_{a}^{w})^{n_w}
(q_{b}^{u})^{l_u}  (q_{b}^{v})^{l_v}
(q_{b}^{w})^{l_w} (q_{c}^{u})^{m_u}  (q_{c}^{v})^{m_v} \nonumber \\
&\times &
 ({\hat u}^{\dagger})^{n_u + l_u + m_u}
 ({\hat v}^{\dagger})^{n_v + l_v + m_v}
 ({\hat w}^{\dagger})^{n_w + l_w }
 |0>_{u} |0>_{v}|0>_{w}, \label{StOp2}
\end{eqnarray}
where the $p'$ $(p'=n',l',m')$ summation indicates all summations
for non-negative numbers $p_u$, $p_v$, and $p_w$  such that  $p_u +
p_v + p_w = p$ $(p =n, l, m) $. When the detector in the $u$ mode
detects a single photon, the state in Eq. \ref{StOp2} can be written
as follows:
\begin{eqnarray}
 (|1>_u)^\dagger |\xi,\beta,\gamma>
  &=& e^{-\frac{1}{2} (|\beta|^2+|\gamma|^2 )} \sum_{n_u + l_u + m_u = 1}  C_{n}(\xi) \beta^{l}\gamma^{m} \frac{\sqrt{n!}}{n_u ! n_v ! n_w !}
   \frac{1}{l_u ! l_v ! l_w !}
\frac{1}{m_u ! m_v! }  \nonumber \\
   & \times &
(q_{a}^{u})^{n_u}  (q_{a}^{v})^{n_v} (q_{a}^{w})^{n_w}
(q_{b}^{u})^{l_u}  (q_{b}^{v})^{l_v}
(q_{b}^{w})^{l_w} (q_{c}^{u})^{m_u}  (q_{c}^{v})^{m_v} \nonumber \\
&\times &
 ({\hat v}^{\dagger})^{n_v + l_v + m_v}
 ({\hat w}^{\dagger})^{n_w + l_w }
|0>_{v}|0>_{w}, \label{StOp2a} \\
& =&  \sum_{ N_v = 0, N_w = 0 } C(N_u =1 , N_v ,  N_w)
    |N_v > |N_w>,  \label{simpleEq}
\end{eqnarray}
where we used new variables  $N_u = n_u + l_u + m_u $,
  $N_v = n_v + l_v + m_v $, and $N_w = n_w + l_w$, and $ |C(N_u =1 , N_v ,
  N_w)|^2$ is the
probability of finding $N_v , N_w $ photons in the output modes $v,
w$ when a detector in the $u$ mode detects a single photon.

\section{Analytic conditions to generate an entangled State}

In order to find the generating condition for the entangled coherent
state, we investigate the coefficient $ C(N_u =1 , N_v = n , N_w = m
)$ in detail.  Let us find the state entangled coherent state by
first finding the analytic solution for generating the state $|0>_v
|\psi>_w $. If  we collected the coefficients for $N_u = n_u + l_u +
m_u = 1$ and $ N_v =  n_v + l_v + m_v =0 $ from the Eq. \ref{StOp2a}
with $n_v = l_v = m_v =0$ and $n_u + l_u + m_u = 1 $, then Eq.
\ref{StOp2a} becomes
\begin{eqnarray}
|0>_v |\psi >_w
  &=& e^{-\frac{1}{2} (|\beta|^2+|\gamma|^2 )} \sum_{N_w = l_w + n_w } C_{n_u + n_w }(\xi) \beta^{l_u + l_w }\gamma^{m_u }
   \frac{\sqrt{(n_u +  n_w )!}}{n_u !  n_w !}
   \frac{1}{l_u ! l_w !}
\frac{1}{m_u ! }  \nonumber \\
   & \times &
(q_{a}^{u})^{n_u}  (q_{a}^{w})^{n_w} (q_{b}^{u})^{l_u}
(q_{b}^{w})^{l_w} (q_{c}^{u})^{m_u}
  ({\hat w}^{\dagger})^{n_w + l_w }
  |0>_{v}|0>_{w}. \label{StOp3}
\end{eqnarray}
$ (n_u, l_u , m_u ) $  set has three cases $(1,0,0), (0,1,0)
,(0,0,1) $ for $n_u + l_u + m_u = 1 $, and so $ |\psi >_w $ in Eq.
\ref{StOp3} can be written
\begin{eqnarray}
 |\psi >_w
  &=& e^{-\frac{1}{2} (|\beta|^2+|\gamma|^2 )} \sum_{N_w = l_w + n_w }
\beta^{ l_w } (q_{a}^{w})^{n_w} (q_{b}^{w})^{l_w}  \frac{1}{l_w !
n_w !} \nonumber \\ & \times &
 [ C_{ 1+n_w } (\xi) q_{a}^{u}
  \sqrt{(1+  n_w )!}   +  C_{n_w }   \sqrt{n_w !} ( \beta  q_{b}^{u}
 +  \gamma q_{c}^{u} )  ]  ({\hat w}^{\dagger})^{n_w + l_w }
  |0>_{w}. \label{StOp4}
\end{eqnarray}

If we set the amplitude of the two modes ($\beta , \gamma $) and the
transmittance of two beam splitters such that
\begin{eqnarray}
  \beta  q_{b}^{u}  + \gamma q_{c}^{u}  = 0 , \label{bgZero}
\end{eqnarray}
using the Eq. \ref{eqOPabc}, the transmittance $ t_2 $ can be found
which satisfy Eq. \ref{bgZero} for a given $ \beta, \gamma, t_1 $
\begin{eqnarray}
  t_2  = \frac{\gamma}{ \sqrt{  \beta^2 + \gamma^2 - \beta^2 t_1 }}.
 \label{t2sol}
\end{eqnarray}
Then  $ |\psi >_w $  can be written
\begin{eqnarray}
 |\psi >_w
  &=& e^{-\frac{1}{2} (|\beta|^2+|\gamma|^2 )} \sum_{N_w = l_w + n_w }
\beta^{ l_w } (q_{a}^{w})^{n_w} (q_{b}^{w})^{l_w}  \frac{1}{l_w !
n_w !} \nonumber \\ & \times &
  C_{ 1+n_w }(\xi) q_{a}^{u}
  \sqrt{(1+  n_w )!}  ({\hat w}^{\dagger})^{n_w + l_w }
  |0>_{w}. \label{StOp5}
\end{eqnarray}
Using $C_{1+n_w}(\xi)$ in Eq. \ref{sV},
\begin{eqnarray}
 |\psi >_w
  &=& e^{-\frac{1}{2} (|\beta|^2+|\gamma|^2 )} \sum_{N_w = l_w + n_w }
\beta^{ l_w } (q_{a}^{w})^{n_w} (q_{b}^{w})^{l_w}  \frac{1}{l_w !
n_w !} \nonumber \\ & \times &
  \frac {\sqrt{(1+ n_w) !}}{\sqrt{\cosh s} \frac{1+ n_w }{2}!}
  (- \frac {1}{2} e^{i \phi} \tanh s )^{\frac{1+ n_w }{2}}
  q_{a}^{u}
  \sqrt{(1+  n_w )!}  ({\hat w}^{\dagger})^{n_w + l_w }
  |0>_{w}, \label{StOp6}
\end{eqnarray}
where $n_w $ should be odd. Then the coefficient for $N_w =l_w +
n_w=0 $ is zero, and the probability that $|\psi >_w $ is in a
vacuum state is zero.  Therefore, $ |\psi>_w$ can not be a coherent
state $(|\alpha>)$.
\begin{figure}[htbp]
\centering
\includegraphics[width=10cm]{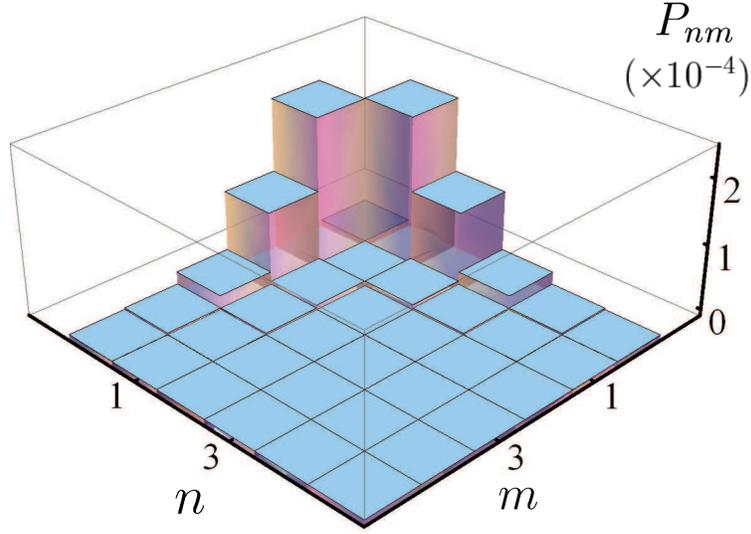}
\caption{ Probability function $P_{nm} $  of the generated states
with  $\alpha = \beta = \frac{1}{2} $,  $s = \frac{1}{2}$ ,
$t_1=0.999 $ and  $ t_2 = 0.999$.} \label{pict999}
\end{figure}
With  integer $k$ such that $ 1+n_w = 2 k $,  and $l_w = N_w - 2k +
1$
\begin{eqnarray}
 |\psi >_w
  &=& e^{-\frac{1}{2} (|\beta|^2+|\gamma|^2 )}\frac{2 q_{a}^{u}}{ \sqrt{\cosh s}} \sum_{N_w =1}
   \sum_{k= 1} ^{\frac{N_w + 1}{2}} (\beta q_{b}^{w})^{N_w}
(\frac{q_{a}^{w}}{ \beta  q_{b}^{w}} )^{ 2 k -1} \nonumber \\ &
\times & \frac{1 }{(N_w -2 k +1) ! (k-1) ! }
  (- \frac {1}{2} e^{i \phi} \tanh s )^{k}
   ({\hat w}^{\dagger})^{N_w }
  |0>_{w}. \label{StOp9}
\end{eqnarray}
Considering the summation over $k$, two terms compete. One of them
is  $ |(\frac{q_{a}^{w}}{ \beta q_{b}^{w} })^2  (- \frac {1}{2} e^{i
\phi} \tanh s )|^k  $, which  decrease as $k$ increases if $
|(\frac{q_{a}^{w}}{ \beta  q_{b}^{w} })^2  (- \frac {1}{2} e^{i
\phi} \tanh s )|<1  $. The other one is $ \frac{1 }{(N_w -2 k +1) !
(k-1) ! } $, which increase as $k$ increases.  So the summation over
$k$ can not turn into a simple analytic form. However, if the
following condition satisfied
\begin{eqnarray}
|(\frac{q_{a}^{w}}{ \beta  q_{b}^{w} })^2  (- \frac {1}{2} e^{i
\phi} \tanh s )| << 1,
 \label{smallv}
\end{eqnarray}
the dominant term in the  $k$ summation is only for  $k=1$, and then
$ |\psi
>_w $ becomes
\begin{eqnarray}
 |\psi >_w
  &\sim & e^{-\frac{1}{2} (|\beta|^2+|\gamma|^2 )}\frac{2 q_{a}^{u}}{ \sqrt{\cosh s}}
   (\frac{q_{a}^{w}}{ \beta q_{b}^{w} }) (- \frac {1}{2} e^{i \phi} \tanh s )
 \nonumber \\ & \times &
     \sum_{N_w =1}
  (\beta q_{b}^{w})^{N_w}
\frac{1 }{(N_w -1) !  }
   ({\hat w}^{\dagger})^{N_w }
  |0>_{w}. \label{StOp10}
\end{eqnarray}

The terms in Eq. \ref{StOp10} are not matched to the coherent state,
but the forms are similar to the photon added coherent state $ {\hat
w}^{\dagger} |(\beta q_{b}^{w}) >_w $.

If we set the same amplitudes of two coherent beams $\beta_0 =
\gamma_0  = \frac{1}{2} $,  $s = \frac{1}{2}$ , and $t_1=0.999 $ ,
 Eq. \ref{t2sol} satisfied with $ t_2 = 0.999$.  These setups give $
|(\frac{q_{a}^{w}}{ \beta  q_{b}^{w} })^2  (- \frac {1}{2} e^{i
\phi} \tanh s )|= 1.85 \times 10^{-3}  $, and then  $ |\psi>_w $ may
be written as Eq. \ref{StOp10}. In Fig. \ref{pict999}, we plot the
probability
\begin{eqnarray}
P_{nm} = | C(N_u =1 , N_v = n ,  N_w = m )|^2.  \label{Pnm}
\end{eqnarray}
calculated from the original Eq. \ref{simpleEq}. The probability
$P_{nm} =0 $, if $n \neq 0$ or $m \neq 0$. Although, we calculated
the analytic form for the  $|0>_v |\psi_1 >_w $ state, the
probability distribution $P_{nm}$ in Fig. \ref{pict999} shows the
existence of another state $|\psi_2 >_v |0>$. We assumed that the
state has a form of the entangled state such as ( $|0>|\psi_1
>_w + |\psi_2 >_v |0> $).

\section{Numerical Conditions to Generate an Entangled Vacuum Evacuated Coherent State.}

\begin{figure}[htbp]
\centering
\includegraphics[width=10cm]{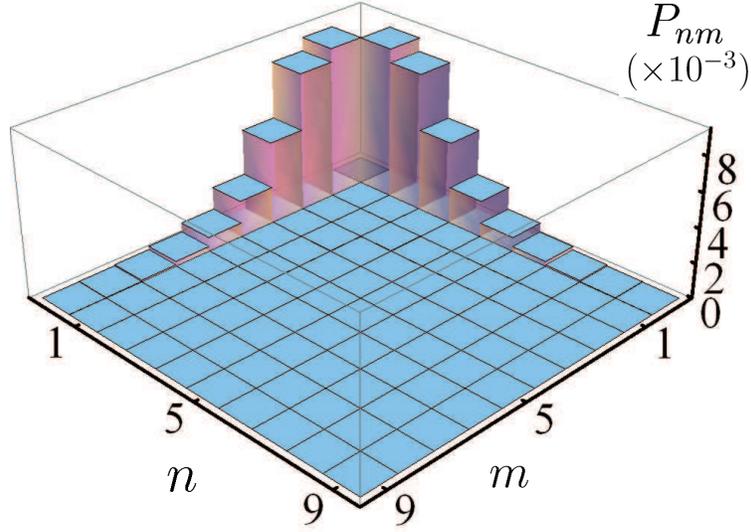}
\caption{Probability function $P_{nm}$  of the generated states with
$\beta  = 0.81  $ .} \label{picA0p81}
\end{figure}

 In order to study the state  ($|0>|\psi_1
>_w + |\psi_2 >_v |0> $) in detail, we release the condition Eq. \ref{smallv},
and we numerically find some entangled quantum
 state by changing parameters with fixed $s$ and $\gamma_0$. Several conditions
 to generate an entangled quantum state
are given in  Table \ref{EC2}.  To generate the state $|\Psi_1 > $ ,
we set the amplitudes of two coherent beams $\beta_0 = 0.813,
\gamma_0 = \frac{1}{2} $, and the squeezing parameter  $s =
\frac{1}{2}$ , and two transmittance of the beam splitters are
$t_1=0.829 $, and $t_2 = 0.740$. With these setups, if we detect a
single photon at the $u$ mode, then we are sure that an entangled
quantum state $|\Psi_1
> $ is generated at the $v$ and $w$ modes.
We define the purity ($Pu $)
\begin{eqnarray}
Pu =  \frac{\sum_{n=0}P_{n0} +\sum_{m=0} P_{0m} }{\sum_{n=0,m=0}
P_{nm}}. \label{purity}
\end{eqnarray}
Then the $|\Psi_1 > $  in Table  \ref{EC2} has $Pu = 99.8 \% $. The
purity $Pu = 99.8 \% $  means that the generated state is $99.8 \%$
described by $(|0>|\psi_1 >_w + |\psi_2 >_v |0>)$, and others can be
represented  $(|n>|\psi_1 >_w + |\psi_2 >_v |m>)$, for nonzero $n,m
$. In Fig. \ref{picA0p81}, we plotted the probability $P_{nm}$ for
$|\Psi_1 > $ in Table \ref{EC2}. We can see that $P_{nm}$ is almost
zero if one of the $(n,m) $ is not zero.

The probability $(Pr) $ to make such event is as follows:
\begin{eqnarray}
Pr =  \sum_{n=0,m=0} P_{nm}, \label{pr}
\end{eqnarray}
then the $Pr$  for the $|\Psi_1 > $  in Table  \ref{EC2} is $ 0.050
$.

In order to find the characteristics of the probability amplitude in
detail, we put the state $|\Psi_1 > $ as follows:
\begin{eqnarray}
|\Psi_1>  =  f (  |0>_v | \alpha>_w^0   - | -\alpha>_v ^0 |0>_w ),
\label{psiw}
\end{eqnarray}
where $|\alpha>^0 ( \equiv |\alpha> - e^{-|\alpha|^2 }|0>)$ is the
vacuum evacuated coherent state. We determine the amplitude $\alpha$
and the scale factor $f$ such that minimizes the error functions
$Er$ as follows:
\begin{eqnarray}
Er = \sum_{n = 1}^9 ( | C(1,n,0) +  f \times co(-\alpha,n) |^2   + |
C(1,0,n) - f \times co(\alpha ,n) |^2), \label{Err}
\end{eqnarray}
where $co(\alpha , n )$ is the coefficient such that the state is in
the number state $|n>$ for the coherent state with amplitude
$\alpha$
\begin{eqnarray}
 co (\alpha , n ) = e^{-|\alpha|^2} \alpha ^n / \sqrt{n!}.
   \label{Co}
 \end{eqnarray}
Note that, Eq. \ref{Err} does not count the $co(\alpha,0)$ and the
summation is limited at the photon number 9.

\begin{figure}[htbp]
\centering
\includegraphics[width=10cm]{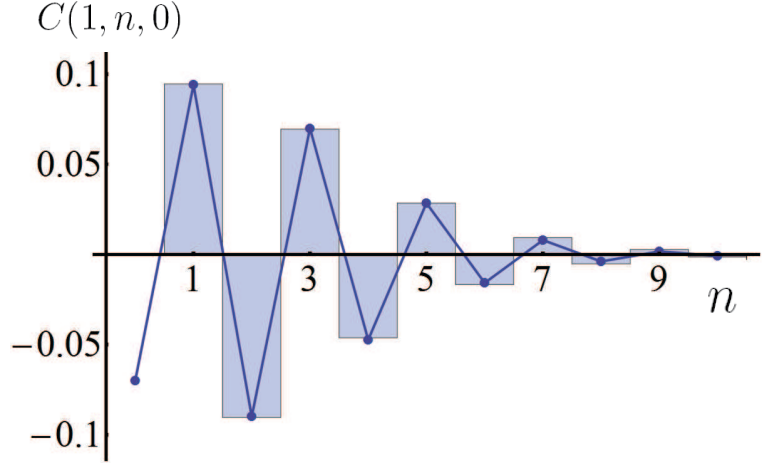}
\caption{The bar chart represents the coefficient $C(1,n,0)$ and the
jointed dots represent  $(- 0.174 \times co(-1.350,n))$ .}
\label{picAlpha}
\end{figure}

\begin{figure}[htbp]
\centering
\includegraphics[width=10cm]{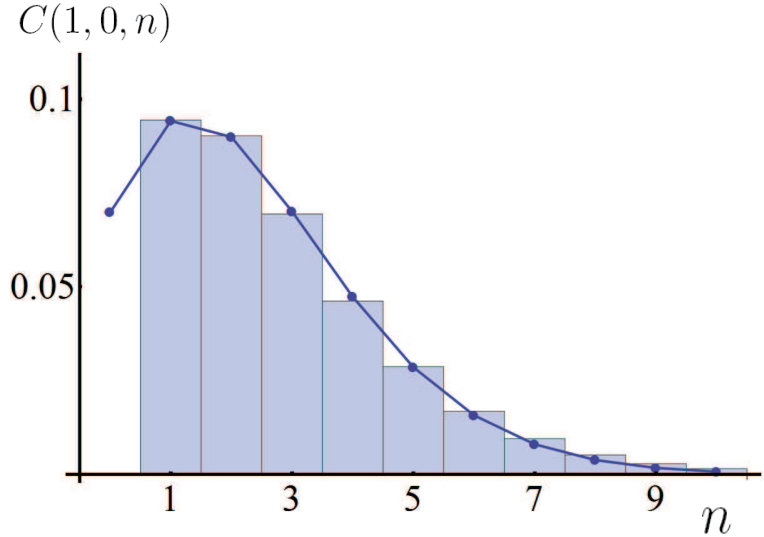}
\caption{The bar chart represents the coefficient $C(1,0,n)$ and the
jointed dots represent  $( 0.174 \times co(1.350,n))$ .}
\label{picBeta}
\end{figure}

In Fig. \ref{picAlpha},  we plotted the coefficient $C(1,n,0)$
defined by Eq. \ref{simpleEq} and $(- f \times co(-\alpha,n))$. The
bar chart represents $C(1,n,0)$ and the joined dots represent $
0.174 \times co(-1.350, n ) $.  We also plotted the coefficient
$C(1,0,n)$ and $f \times co(\alpha,n)$ in Fig. \ref{picBeta}.  The
joined dots in Fig. \ref{picBeta} were calculated from the
coefficient $ 0.174 \times co (1.350,n) $.  The total error sum is
about  $1.6 \times 10^{-5}$.  If we represents the state
$|\alpha>_w^0$ as a in number state, the coefficient is the same as
that of the coherent state only excepting only the coefficient of
the vacuum state.


 The probability amplitude $P_{nm}$ is sensitive to the phase
 relation among three input beams. In addition to the phase
 relations among three beams, the phase shift in Eqs. \ref{BSab}-\ref{BScd} at two beam splitters
 should be counted. The total phase factors are linear functions of
 the relative phases among the three input beams and the phase shifts at two beam
 splitters  \cite{youn13}, so if we scan the relative phases $\theta$ and $\phi$, we
 can obtain the phase-dependent probability amplitude $P_{nm}$. For
 simple notation, we set all the phase shifts at the beam splitters to $0$.

 If we changed the phase $\phi$ from $\pi$ to $0$, the probability
 amplitude $P_{nm}$ is not an entangled vacuum evacuated coherent
 state any more. If the squeezed sate is divided by a beam splitter, the entanglement
  between the two outputs is very sensitive to the phase shifts
  \cite{MsKimQ}.
  In actual experiments, it's difficult to control the phase
  shift of the beam splitter. However, in our scheme, the total phase
  shifts including the phase shifts at the beam splitter  can be
  scanned  by changing the phase differences among the  three input
  beams. The entangled vacuum evacuated coherent state can be
  obtained using only the phase differences $\phi =\pi $ and $ \theta = \pi$.

\begin{figure}[htbp]
\centering
\includegraphics[width=10cm]{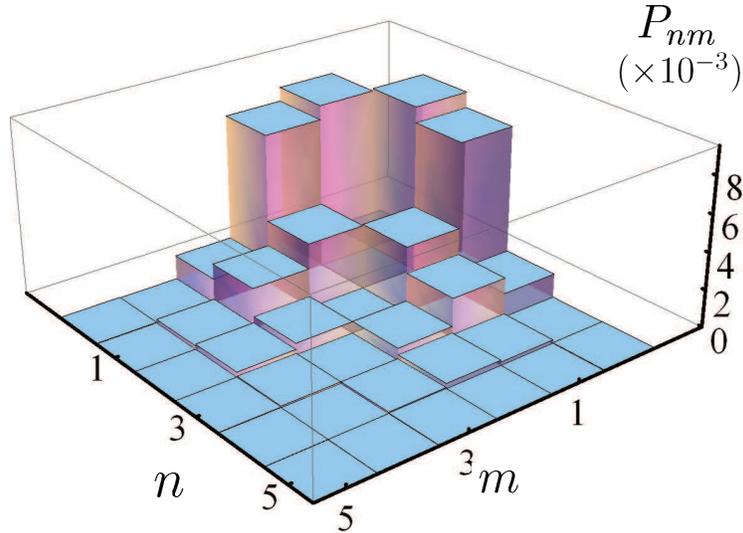}
\caption{Probability function $P_{nm}$  of the generated states with
$\beta  = 0.81  $ , $ s = \frac{1}{2}$ , $ \phi = 0$.}
\label{picA0p81pZero}
\end{figure}

\begin{figure}[htbp]
\centering
\includegraphics[width=10cm]{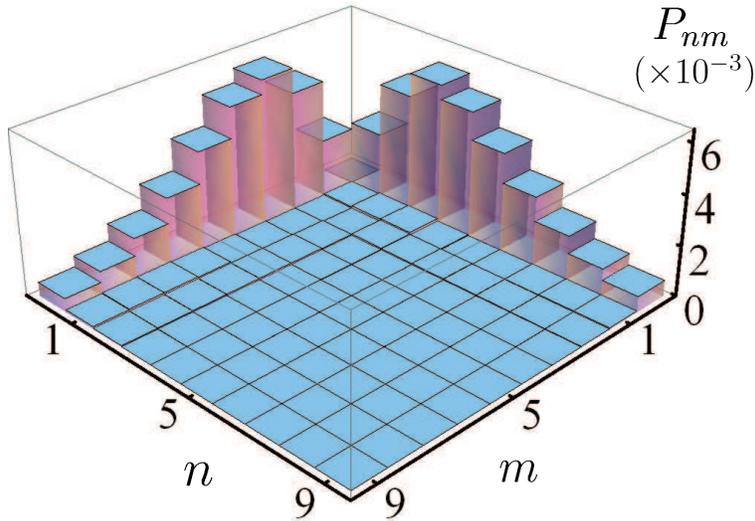}
\caption{Probability function $P_{nm} $  of the generated states
with  $\beta  = 1.302 $ .} \label{picA1p30}
\end{figure}

We showed several conditions to generate an entangled vacuum
evacuated coherent state for bigger amplitudes in the Table
\ref{EC2}. With the input amplitudes $\gamma_0 = \frac{1}{2} $, $ s
= \frac{1}{2}$, we can obtain  the amplitude of the entangled vacuum
evacuated coherent state amplitudes of  $\alpha= 1.591$ and $\alpha
= 1.819$. If we add the amplitude of the squeezed vacuum $s=3/4$,
the amplitudes of the entangled vacuum evacuated coherent state
become $\alpha =1.990$ and $\alpha = 2.160$. We plotted the
probability amplitude $P_{nm}$ for $|\Psi_4 > =  (|0>|1.990>^0 -
|-1.990>^0 |0>)$  in Fig.\ref{picA1p30}.

The probability $Pr$  to generate $|\Psi_1>$ ~ $|\Psi_5> $ is around
$5 \%$, and the purity $Pu$, is greater than $96 \%$. The err sum
$Er$ is increased as $\alpha $ is increased. The main reason is
caused by our calculation limit. In Fig. \ref{picA1p30}, the
probability amplitude $P_{09}$  and $P_{90}$ are not zero. We only
calculated the Error sum to the photon number $9$ in Eq. \ref{Err}.
The main reason we terminated the photon number at $9$ is to make
the computational and memory size burdens reasonable. Although we
terminated the coefficients at $9$, the coefficient is the exact one
up to photon number  $9$ in our calculation.

\begin{table}
\centering \caption{Conditions for generating the vacuum evacuated
entangled coherent state. ($\phi = \pi, ~\theta = \pi $)}
\label{EC2}
\begin{tabular}{c||c|c|c|c|c|c|c|c|c|r}
\hline state  & $s$  & $ \beta_0$ & $\gamma_0$ & $t_1$ & $t_2$ &
$Pr$ & $Pu$ (\%) & $\alpha$ & $f$  & $Er (\times 10^{-5})$  \\
\hline
 $ |\Psi_1 >  $  & $1\over 2 $ & 0.813 &  $1 \over 2$  & 0.829  & 0.740
  &  0.050  & 99.8 & 1.350 & 0.174 & 0.16 \\ \hline
  $ |\Psi_2 >  $ & $1 \over 2 $ & 1.040 &
$1\over 2$  & 0.780  & 0.609  &  0.041 &99.3  & 1.591 & 0.148 & 0.33  \\
\hline $ | \Psi_3 >$  & $1 \over 2 $ & 1.292 &
$1\over 2$  & 0.744  & 0.502  &  0.029 & 97.1 & 1.819 & 0.122 & 4.25 \\
\hline $ |\Psi_4 > $ & $3 \over 4 $ & 1.302 &  $1 \over 2$  & 0.729
& 0.491  & 0.059  &97.1  & 1.990 & 0.171 & 24.8 \\ \hline
 $ |\Psi_5 > $ & $3 \over 4 $ & 1.498 &  $1 \over 2$  & 0.710 & 0.428
 &   0.046 &96.2  & 2.160 & 0.150  &  18.2  \\ \hline
         \hline
\end{tabular}
\end{table}
%

%
%
%
%

%
%
%
%
%
\section{Discussion}

   The generation of entangled states through the use of a squeezed
light source and conditional measurements has been extensively
studied both theoretically and experimentally. Our work is to
demonstrate how to generate an entangled state  with a squeezed
light, two coherent lights, two beam splitters and single photon
counters. The two beam splitters and the two coherent beams give us
a degrees of freedom with which  to control the output in a highly
nonclassical manner. We characterize  the two outputs from the three
input beams with the  detection of single photon.

   Entangled quantum states are indispensable for quantum information science.
   Furthermore,
   entangled coherent states  have great advantages for applications. In
 our new scheme to generate an entangled quantum state, we showed the
 possibility  of generating a vacuum evacuated coherent state that is
 entangled with vacuum state. We find the analytic conditions for
 which an entangled quantum state is generated at
 the output port $w$ and $v$ when a single photon is detected at the output port
 $u$. With the same amplitudes $\beta_0 = \gamma_0 = s = \frac{1}{2}$,
 and $t_1 = t_2 = 0.999 $, an entangled quantum state can be generated.

 If we changed the amplitude $\beta_0 = 0.813$, and
the two transmittances of the beam splitters are $t_1=0.829 $, and
$t_2 = 0.740$, we can obtain the entangled quantum state
$(|0>|\psi_w> + |\psi_v>|0>)$.  The state $|\psi_w >$ is a vacuum
evacuated coherent state $(|\alpha>^0  = |\alpha> - e^{-|\alpha|^2
}|0> )$ with $\alpha = 1.350 $, and $| \psi_v> = - |\alpha = -1.350
> ^0$.  The error sum is less than $10^{-5}$. We can obtain  a large
amplitude of the entangled vacuum evacuated coherent state up to
$\alpha= 2.160$ with the amplitude of the squeezed vacuum
$s=\frac{3}{4}$. Although the error sum $Er$ is increased as $\alpha
$ is increased, the main reason is simply caused by calculation
setting limit we set for fast calculation.

 With the explicit form, the  probability
amplitude for an output state is a function of the transmittances of
the two beam splitters and the amplitudes and the relative phases of
the three input beams. The probabilities are calculated when the a
single photon is detected.  We have included all of the coefficients
of the input beams from zero to 16 of the number representations for
the three input states and use the coefficient up to $9$.

   Considering applicability to actual experiments \cite{Dong2014},
 if we use the input beam as
 pulsed light with a repetition rate of
100 MHz, then a generation probability $Pr$  of  $10^{-3}$ results
in  $10^6$ signals per second.

   In actual experiments, a entanglement can be reduced
    as a result of  experimental imperfections, such as mode matching and non-unity quantum
efficiency. We assumed perfect temporal and spatial mode matching
among the three input beams. These assumptions also guaranteed that
the spatial and the temporal mode properties of the entangled states
generated in our scheme are well defined by the input states and
that the modes of the two coherent states and the squeezed vacuum
could be precisely controlled by adjusting the pump beam used to
produce the squeezed states. We expect high-purity spatial and
temporal modes of the entangled state. A large amplitude entangled
vacuum evacuated coherent state can be used to study the quantum
nature of the world,  and it is  a key element in quantum
technology.

\acknowledgements

 This study was supported by the Basic Science
Research Program through the National Research Foundation of Korea
(NRF) funded by the Ministry of Education, Science and Technology
(NRF-2014R1A1A2055454)

%
%

\end{document}